\documentclass[twocolumn,showpacs,preprintnumbers,amsmath,amssymb]{revtex4}                    

\usepackage{graphicx}
\usepackage{dcolumn}
\usepackage{bm}
\usepackage{tabularx}
\usepackage{amsmath}

\begin{document}

\title{Anomalous reduction of the Lorenz ratio at the quantum critical point in YbAgGe}

\author{J. K. Dong,$^1$ Y. Tokiwa,$^1$  S. L. Bud'ko,$^2$ P. C. Canfield,$^2$ P. Gegenwart$^{1}$}

\affiliation{$^1$I. Physikalisches Institut, Georg-August-Universit\"{a}t G\"{o}ttingen, 37077 G\"{o}ttingen, Germany\\
$^2$Ames Laboratory and Department of Physics and Astronomy, Iowa State University, Ames, Iowa 50011, USA}

\date{\today}

\begin{abstract}
We report measurements of the electrical and thermal transport on the hexagonal heavy-fermion metal YbAgGe for temperatures $T\geq$ 40 mK and in magnetic fields $H\parallel ab$ up to 14 T. This distorted Kagome-lattice system displays a series of magnetic states and a quantum critical point at $H_c =4.5$~T. The Lorenz ratio $L(T)/L_0$ displays a marked reduction only close to $H_c$. A $T$-linear contribution below 120 mK, present at all different fields, allows to extrapolate the Lorenz ratio towards $T=0$. At the critical field this yields $L/L_0=0.92\pm 0.03$, suggesting a violation of the Wiedemann-Franz law due to strong inelastic scattering.
\end{abstract}

\maketitle

The realization of quantum critical points (QCPs), i.e. continuous transitions at $T=0$ driven by the change of a non-thermal parameter, has led to the discovery of interesting new phenomena like non-Fermi liquid (NFL) behavior and unconventional superconductivity in recent years~\cite{Stewart,Lohneysen1,Gegenwart,Lohneysen}. Heavy fermion (HF) systems consist of unstable local magnetic moments (often from Ce-,Yb- or U-atoms) in metallic environment and are prototype materials for the investigation of QCPs. Their ground state is determined by the competition between the Kondo- and Ruderman-Kittel-Kasuya-Yosida (RKKY) interaction~\cite{Doniach}. As a result of their sensitivity to small changes in pressure, composition or magnetic field, HF compounds represent an ideal playground for the controlled realization of novel quantum phases~\cite{Petrovic,Mathur}.

Two different scenarios were proposed to describe QCPs
in HF metals: the spin-density-wave (SDW) and the Kondo-breakdown scenario \cite{Hertz,Coleman,Si2001}. A clear distinction between them concerns the role of Kondo singlets when the system is tuned from the paramagnetic side towards the QCP \cite{Si}. In the former scenario, the Kondo singlets survive across the QCP and the magnetic order is of SDW type. This restricts the
critical fluctuations to the vicinity of "hot spots", while the main
part of the Fermi surface remains intact. In striking
contrast, the entire Fermi surface is influenced in the latter case
and the heavy Landau quasiparticles break up due to the
disintegration of the Kondo entanglement \cite{Si}.

Electrical and thermal transport measurements are well established tools to investigate the nature of QCPs~\cite{Kim,Pfau,Machida}. The ultra low-temperature thermal conductivity $\kappa$ of a metallic sample is
related to its electrical conductivity $\sigma$ through the Wiedemann-Franz (WF) law $\kappa/\sigma T = \pi^2 k_B^2/3e^2 = L_0$. Since the WF law is independent of the quasiparticle mass it can be viewed as a touchstone to verify whether Landau quasiparticles survive at the QCP. The WF law is strictly valid at $T=0$ K, if the following necessary conditions are fulfilled: (i) the charge and energy must be carried by the same fermions and (ii) scattering is fully elastic~\cite{Tanatar1}. If for example in a one-dimensional (1D) material charge neutral spinon
excitations are present, the Lorenz ratio $L(T)/L_0$ with
$L(T)=\kappa/\sigma T$ can strongly exceed $1$ in the
zero-temperature limit. In 3D metals, a violation of the WF law may be expected at the Kondo-breakdown QCP where well-defined heavy quasiparticles disintegrate~\cite{Kim,Pfau}.

Experimentally, a violation of the WF law was
reported in the two prototype HF metals CeCoIn$_5$ and YbRh$_2$Si$_2$~\cite{Pfau,Machida,Tanatar1,Tanatar2}. In tetragonal CeCoIn$_5$, the WF law was found to be violated with $L/L_0 \approx 0.8$ at the field-induced QCP when the current flows perpendicular to
the layers, while it turns out to hold for the in-plane transport
\cite{Tanatar1,Tanatar2}. The violation of the WF law results from linear extrapolation of the thermal and electrical resistivity towards absolute zero, which, however, has been questioned subsequently~\cite{Smith}. For YbRh$_2$Si$_2$, two recent studies on the thermal and electrical conductivity near the field-tuned QCP have found rather similar experimental data obtained for fields either perpendicular~\cite{Pfau} (down to 25 mK) or parallel~\cite{Machida} (down to 40 mK) to the tetragonal $c$-direction at the respective critical fields. However, these studies strongly differ in the extrapolation of the thermal resistance at the critical field for $T\rightarrow 0$. While the former study attributed an observed downturn of the thermal resistance below 0.1~K to overdamped magnons, the latter one treated the same downturn as signature of the quasiparticle formation. Exclusion or inclusion of this feature to the $T\rightarrow 0$ extrapolation leads to either a violation ($L/L_0 \approx 0.9$)~\cite{Pfau} or verification~\cite{Machida} of the WF law, respectively. It is therefore badly needed to investigate the validity of the WF law in further quantum critical HF metals.

\begin{figure}
\includegraphics[clip,width=1\columnwidth]{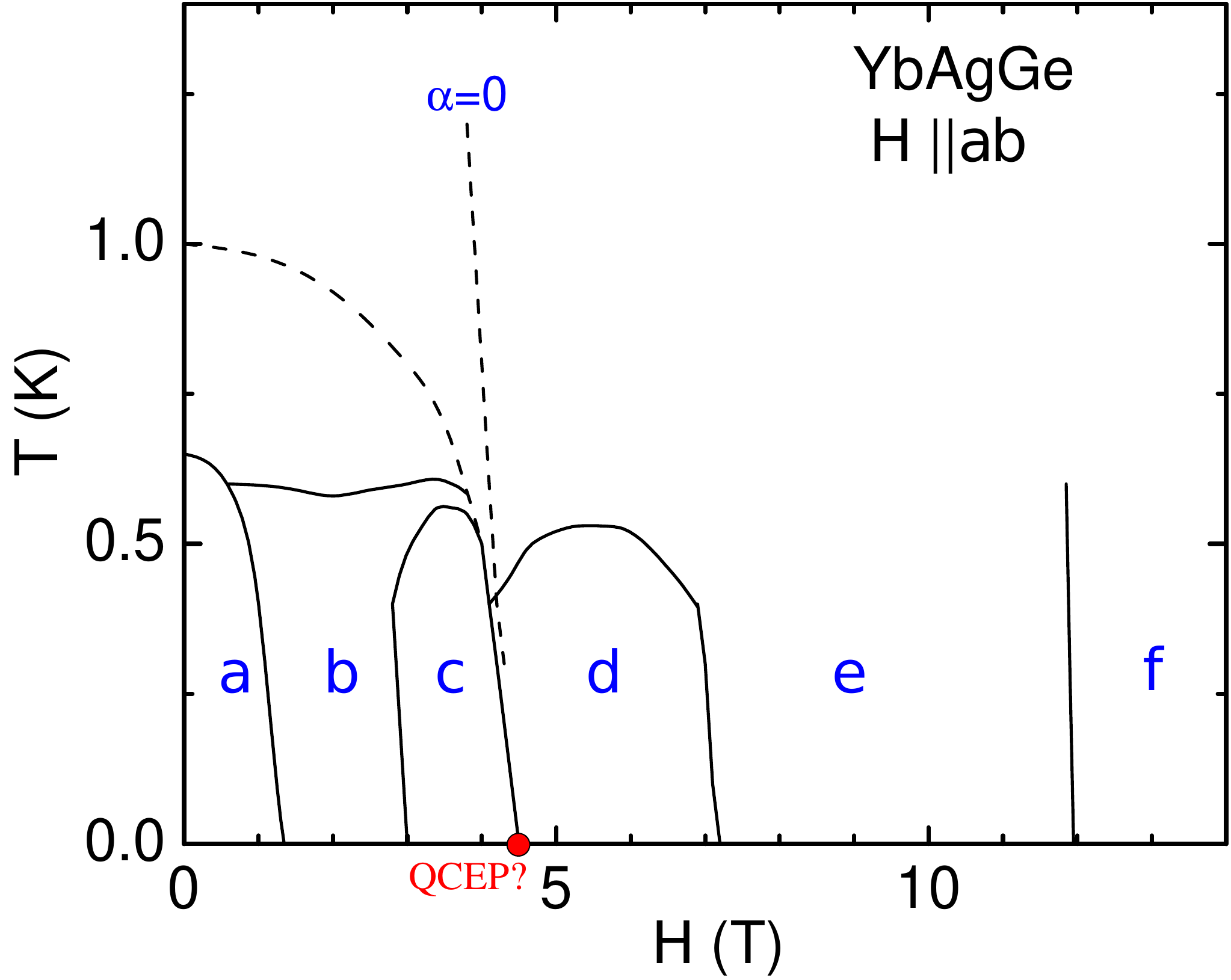}
\caption{(Color online). Schematic $T$-$H$ phase diagram of YbAgGe for $H\parallel ab$, summarizing earlier measurements
\cite{Canfield2011}. Below 4.5 T, commensurate AF order forms in phase $a$, then changes to incommensurate AF order in phase $b$, and finally returns back to commensurate AF order in $c$~\cite{Fak,Fak06}. A sign change of the thermal expansion coefficient ($\alpha$) appears near the border between the regions $c$ and $d$. Electrical resistivity $\rho(T)$ varies as $T^{n}$ with $n \simeq 1$
in region $d$, $1<n<2$ in $e$ and $n=2$ in region $f$~\cite{Canfield2004}. The red circle indicates the position of the presumed quantum critical end point (QCEP) at $H_c=4.5$~T.}
\end{figure}

In this paper, we have performed electrical resistivity and thermal conductivity measurements on the HF metal YbAgGe in order to investigate the quantum critical behavior and the validity of the WF law. Yb ions in YbAgGe with hexagonal ZrNiAl structure form a distorted Kagome lattice, where the magnetic interactions between the first nearest neighbors are fully frustrated. In zero magnetic field, the specific heat displays a broad maximum at 1~K and subsequent sharp antiferromagnetic phase transition at 0.7~K~\cite{Canfield2004}. The highly complex phase diagram in magnetic field, as shown in Fig.\,1~\cite{Canfield2011}, implies the importance of magnetic frustration. Interestingly, the isostructural CePdAl exhibits an unusual partial ordering, with 2/3 of Ce-4$f$ moments ordered and the remaining 1/3 in a paramagnetic state, due to  frustrated interactions \cite{Nishiyama}. For YbAgGe, neutron scattering studies have revealed that the $a$ and $c$ phases are ordered with a commensurate wave vector (1/3, 0, 1/3) while the $b$ phase displays an incommensurate ordering vector (0, 0, 0.325)~\cite{Fak,Fak06}. The microscopic nature of all the other phases is still unknown.

At the critical field separating the $c$ and $d$ phases, $H_c$ = 4.5\,T, the thermal and the magnetic Gr\"{u}neisen ratios diverge as the temperature is lowered, indicating quantum criticality~\cite{Canfield2011,Yoshi2012}. The positions of a cross-over feature in the Hall effect and a sign change of thermal expansion ($\alpha$=0) also extrapolate for $T\rightarrow 0$ to 4.5\,T, at which field a metamagnetic-like signature of the magnetization has been found~\cite{Yoshi}. These recent observations led to a speculation that a new type of quantum critical end point (QCEP) is caused by a spin-flop transition of local moments~\cite{Canfield2011,Yoshi2012}. This is different to the case of itinerant moment metamagnetism as observed in Sr$_3$Ru$_2$O$_7$ and CeRu$_2$Si$_2$~\cite{Grigera,Weickert}. A further field-induced QCP in YbAgGe has been proposed at $H \simeq 7.2$ T, where both the specific heat $C/T$ and thermopower $S/T$ display a logarithmic divergence towards low temperatures~\cite{Canfield2011}. Our measurements, discussed below, indicate that the WF law is satisfied in this NFL region e of the phase diagram. Remarkably, however, we observe a marked suppression of Lorenz ratio $L(T)/L_{0}$ at $H = 4.5$ T. Extrapolation to $T \rightarrow 0$ suggests a moderate violation of the WF law near this critical field.

\begin{figure}
\includegraphics[clip,width=1\columnwidth]{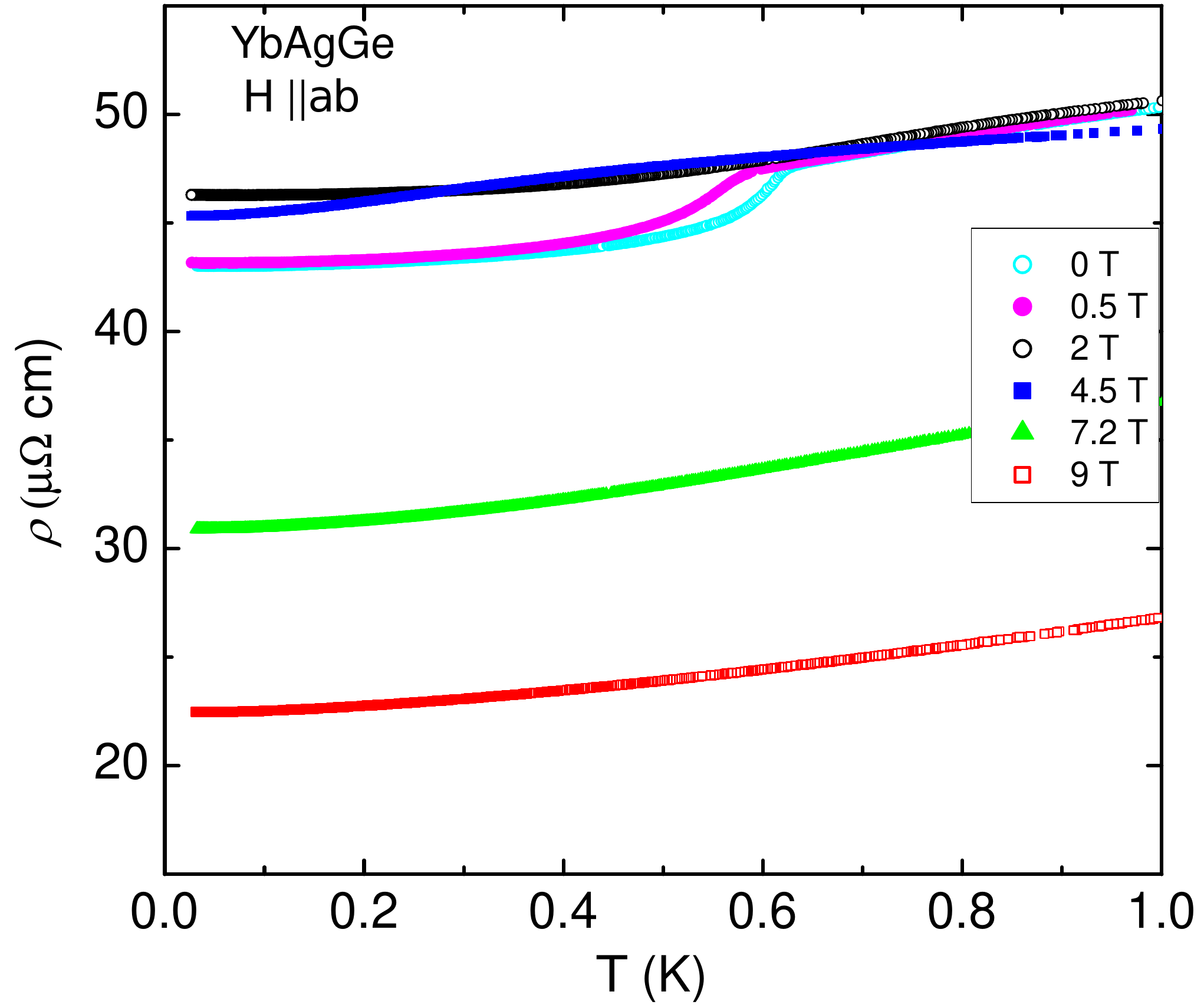}
\caption{(Color online). Temperature dependence of the electrical resistivity $\rho(T)$ of YbAgGe along the $c$-direction at various magnetic fields $H \parallel ab$.}
\end{figure}

The heat and charge transport of YbAgGe were measured parallel to the hexagonal $c$-axis at various magnetic fields applied along the $ab$ plane. Details are provided in the Supplemental Material (SM)~\cite{SM}. Fig. 2 displays the temperature dependence of the electrical resistivity down to 30 mK at selected magnetic fields (further data see SM). The sample used in the present work exhibits a residual resistivity $\rho_0 = 43~\mu\Omega$cm at zero magnetic field. However, this relatively high value is not solely related to structural disorder, since the residual resistivity is strongly reduced by magnetic field and amounts to $\rho_0=19~\mu\Omega$cm, corresponding to a residual resistivity ratio of ${\rm RRR}=5.9$, at 14~T~\cite{SM}. In zero field, the rapid drop of $\rho(T)$ at 0.65 K confirms the first order transition which leads to a clear hysteresis in resistivity and magnetization measurements \cite{Niklowitz,Yoshi}. In low fields, this transition is suppressed to lower temperatures and finally disappears at $H=2$~T. At 4.5 T, a weak shoulder-like
anomaly is found around 0.4 K. At higher fields, the electrical
resistivity can be described by $\rho (T) = \rho_0 + AT^n$ with the exponent $n \simeq 1.60$ and 1.58 for $H=7.2$ and 9 T, respectively. The NFL region thus largely exceeds the proposed QCP at 7.2~T in agreement with the phase diagram constructed by earlier resistivity measurements~\cite{Canfield05}. This may be related to the  negative low-temperature magnetoresistance $d\rho/dH<0$ in this field regime, which saturates only beyond 10~T (see SM~\cite{SM}).

\begin{figure}
\includegraphics[clip,width=1\columnwidth]{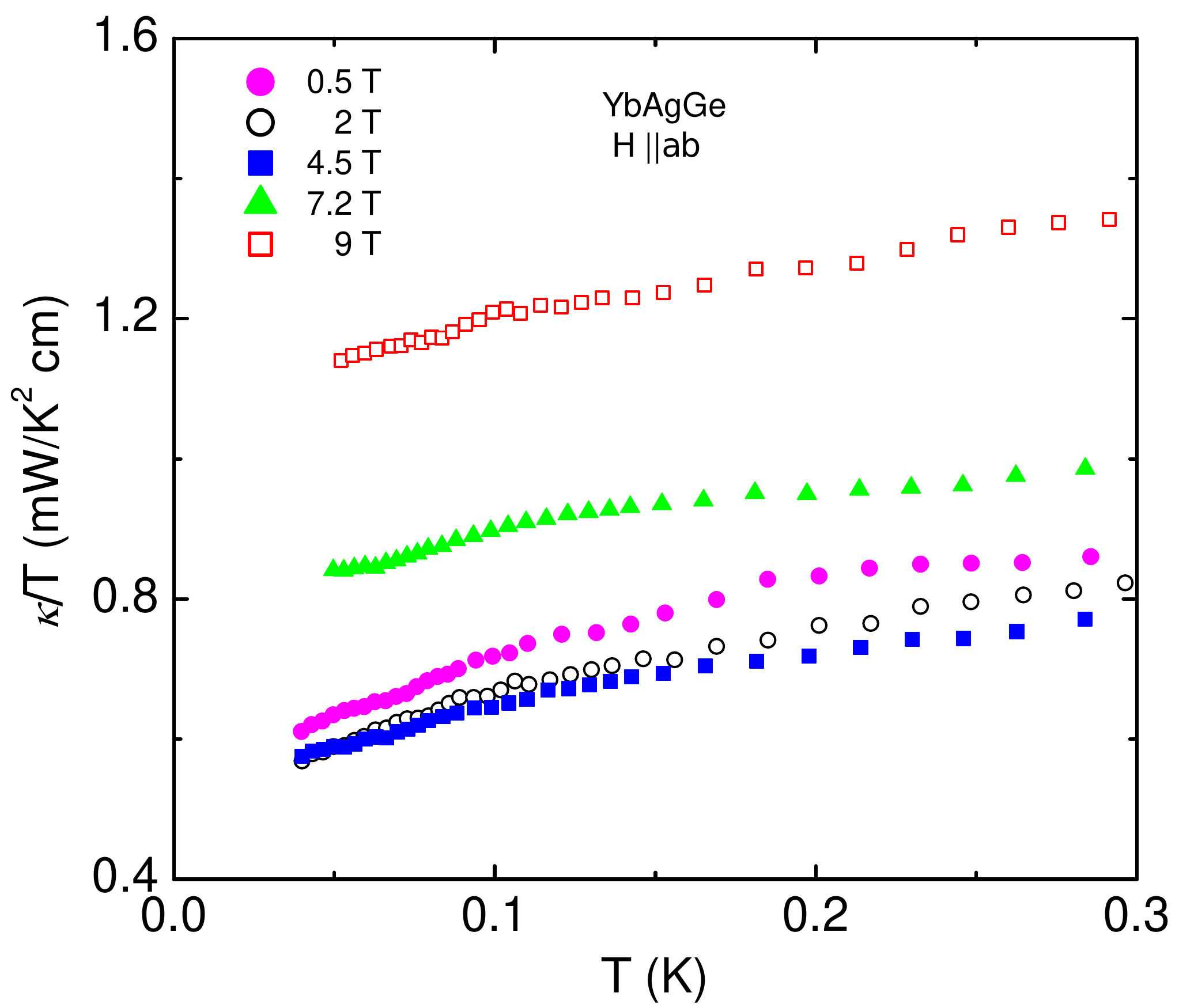}
\caption{(Color online). Temperature dependence of the thermal conductivity divided by temperature $\kappa(T)/T$ of YbAgGe at different magnetic fields.}
\end{figure}

The thermal conductivity $\kappa(T)/T$ of YbAgGe as a function of temperature is displayed in Fig. 3, at temperatures $T \leq$ 0.3 K and different magnetic fields up to 9 T. Further data are shown in SM~\cite{SM}. In general, heavy quasiparticles, phonons and magnons could transport energy in YbAgGe. We plot $\kappa(T)/T$, since Bosonic contributions from phonons and magnons die out at $T\rightarrow 0$, giving rise to a finite intercept due to electrons. Upon cooling to low temperatures, $\kappa(T)/T$ steadily declines in all fields and no indication of saturation is observed. Unlike in YbRh$_2$Si$_2$~\cite{Pfau,Machida} and CeCoIn$_5$~\cite{Tanatar2}, where the total thermal conductivity is dominated by the electronic transport and thus $\kappa(T)/T$ increases with decreasing temperature, the Bosonic contributions in YbAgGe cannot be neglected. The relatively high residual resistivity in YbAgGe strongly reduces the electronic thermal conductivity. This is also reflected in the convex shape of $\kappa(T)/T$ in particular at low fields, which makes it unfeasible to estimate the phonon contribution to thermal conductivity by power-law fitting with reasonable exponent. Therefore it is impossible to analyze the temperature dependence of the electronic thermal resistivity, as done for CeCoIn$_5$ and YbRh$_2$Si$_2$. By contrast, we focus on the total thermal conductivity and the Lorenz ratio at very low temperatures.
The overall behavior presented in Fig. 3, indicates that $\kappa(T)/T$ is gradually suppressed in low fields and reaches a minimum at  $H = 4.5$~T. At higher fields, for $H$ = 7.2 and 9 T, $\kappa(T)/T$ becomes strongly enhanced, which corresponds to the pronounced reduction of the electrical resistivity in this field regime.

Using the electrical resistivity $\rho (T)$ and thermal conductivity $\kappa (T)$ we can calculate $L(T)$ = $\kappa$/$\sigma T$. The temperature dependence of the normalized Lorenz ratio $L(T)/L_{0} =
(\kappa/T)\cdot\rho/L_{0}$ is displayed in Fig. 4. We are interested in the zero-temperature limit. For this purpose it is sufficient to calculate $L(T)$ without subtraction of the unknown additional phonon and magnon contributions, since both will disappear as T $\rightarrow$ 0. We turn our attention to very low temperatures ($T \leq$ 0.15 K) at which the Bosonic contributions gradually die out. In our accessible temperature range above 40 mK, the Lorenz ratio $L(T)/L_{0}>$ 1. Again this behavior, which is caused by the relatively high $\rho_0$ in YbAgGe, is in sharp contrast to the observations in the very clean HF systems CeCoIn$_5$~\cite{Tanatar1,Tanatar2} and YbRh$_2$Si$_2$~\cite{Pfau,Machida} for which $L(T)/L_0<1$ at finite temperatures below 1~K. Empirically, we observe a linear temperature dependence of $L(T)/L_0$ for all investigated fields with no trace of a downturn to lowest temperatures (see SM for further data at additional fields and towards higher temperatures~\cite{SM}). Although we cannot justify this linear temperature dependence, which arises from the inclusion of Bosonic contributions to $\kappa(T)$, it nevertheless allows us to obtain the 0 K values by simple linear extrapolation. At $H \leq 3.5$~T, the Lorenz ratio is weakly field dependent, while we observe a drastic reduction of the low-temperature value at the critical field $H_c =4.5$~T. Upon further increasing the field, it increases again.

\begin{figure}
\includegraphics[clip,width=1\columnwidth]{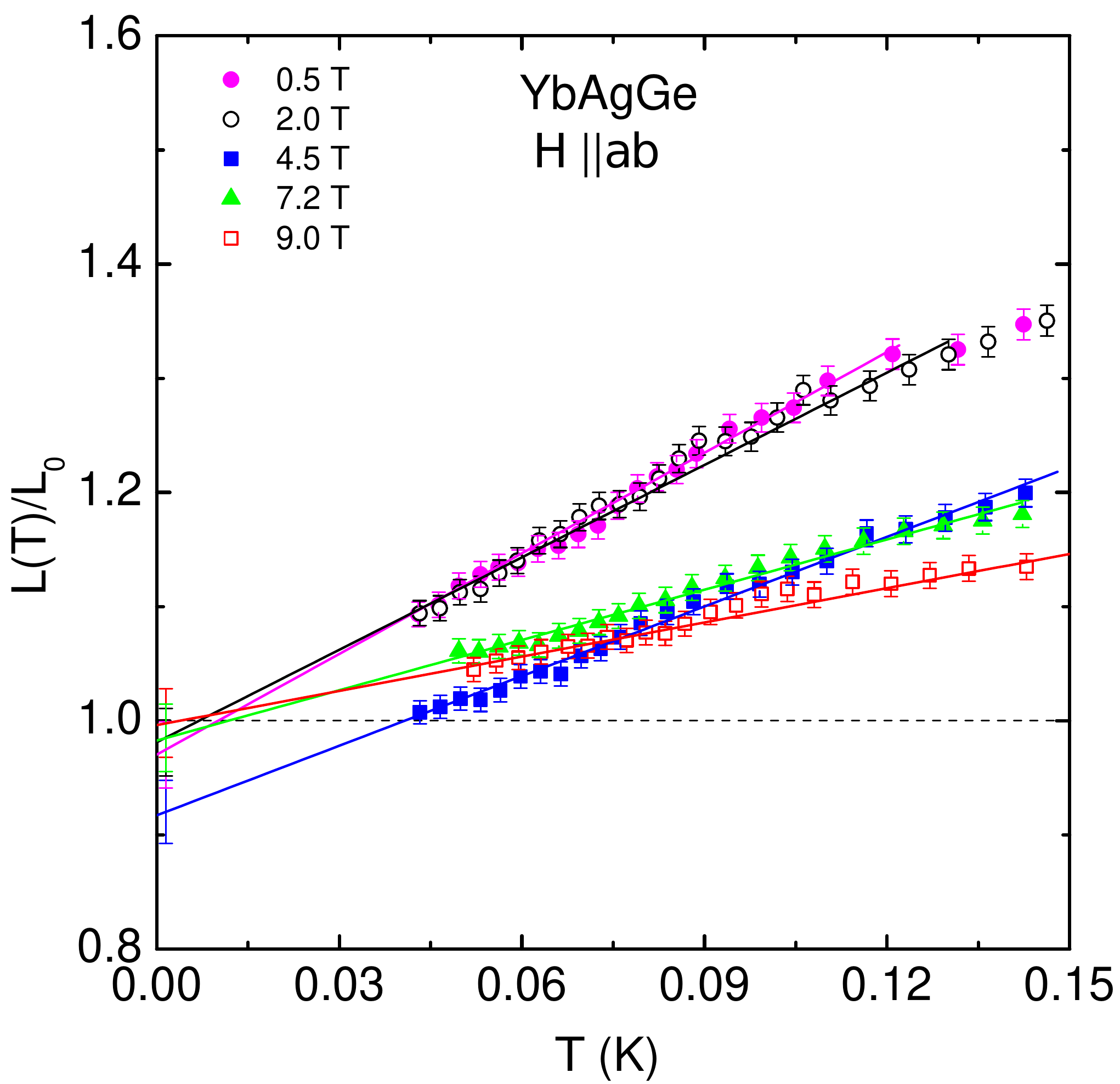}
\caption{(Color online). Temperature dependence of the normalized Lorenz ratio $L(T)/L_{0}$ = $\kappa\rho/(L_{0}T)$ for various different magnetic fields. The solid lines indicate linear fits to the data. Error bars at data points represent the statistical scattering while those at the lines near absolute zero temperature indicate the systematic error for the extrapolated Lorenz ratio arising from the finite width of the contacts.}
\end{figure}

While the depression of $L(50$~mK$)$ clearly indicates anomalous behavior near the QCP, we now turn to the extrapolation of the linear $L(T)$ dependence found at all different fields (see also SM~\cite{SM}) towards $T\rightarrow 0$. It reveals a value significantly below 1 only at the critical field with $L(T\rightarrow 0)/L_0=0.92\pm 0.03$. The observed reduction of $L(T)$ at $H_c$ therefore suggests a violation of the WF law. On the other hand,
for $H = 7.2$ and 9 T, there is a clear change of the slope of $L(T)$/$L_{0}$ and the value of $L(T \rightarrow 0)/L_{0}$ extrapolates to $1.00\pm0.03$ proving that the WF law holds in this extended NFL region of the phase diagram.

The electrical resistivity $\rho(T)$ of YbAgGe is almost temperature independent below 0.1 K in all fields. Therefore, the {\it electronic} thermal conductivity $\kappa_e/T$ should be constant, as observed e.g. in CeNi$_2$Ge$_2$ \cite{Kambe}. According to the relation $\kappa/\sigma T = L(T)$, the slope of the Lorenz ratio reflects the magnitude of the Bosonic thermal conductivity ($\kappa_{\rm ph}$ and $\kappa_{\rm mag}$) due to phonons and magnons. The remarkable depression of the slope in $L(T)/L_0$ at higher fields ($H$ = 7.2 and 9 T) is probably linked to the loss of magnons due to the suppression of the AF order since phonons are unaffected by the magnetic field.

At $H_c = 4.5$~T, the extrapolated $T = 0$ value of $L(T)/L_{0}$ is less than 1, suggesting that the thermal transport ability is significantly depressed by additional scattering sources. Let us return our attention to the raw data of the electrical resistivity $\rho (T)$ and thermal conductivity $\kappa(T)/T$ at $H_c$ in more detail. Compared to the data at $H$ = 2 T, $\rho$(4.5 T) is significantly reduced below 0.2 K, leading to an enhancement of $\sigma$.
However, the thermal conductivity $\kappa$(4.5 T)$/T$ does not show a corresponding increase. The disparate field response of $\sigma$ and $\kappa/T$ therefore leads to the distinct suppression of $L(T)/L_0$ at the critical field 4.5~T compared to all other fields. This experimental observation is independent on the empirical $T\rightarrow 0$ extrapolation and indicates anomalous behavior near the QCP.

It is interesting to compare YbAgGe with the itinerant metamagnets Sr$_3$Ru$_2$O$_7$~\cite{Grigera} and CeRu$_2$Si$_2$~\cite{Flouquet}. For both systems thermodynamic indications for quantum criticality in accordance with the itinerant Hertz-Millis scenario have been found~\cite{Gegenwart06,Weickert}, although the approach of the QCEP is cut-off by the formation of a nematic phase in the former and crossover to Landau Fermi liquid behavior in the latter case. Most interestingly, both materials obey the Wiedemann-Franz law within 5$\%$ error at their quantum critical regions~\cite{Ronning,Pfau1}. The most significant difference between the QCEPs in YbAgGe and these two materials concerns the nature of the magnetic moments. In the above mentioned itinerant systems the quasiparticles stays intact~\cite{JF,Weickert}. Relatedly the WF law is obeyed. For YbAgGe, on the other hand, the metamagnetic behavior is more local in nature, resulting in strong critical fluctuations that may lead to a destruction of the quasiparticles at the entire Fermi surface. At $H \simeq 4.5$~T, previous transport measurements have found an abrupt jump in the field-dependence of the Hall coefficient~\cite{Canfield2005} and sign change of the isothermal thermopower $S(H)$ at low temperatures~\cite{Canfield2010}. These observations indeed suggest a significant change of the Fermi surface at this field.

Another distinct difference of YbAgGe to Sr$_3$Ru$_2$O$_7$ and CeRu$_2$Si$_2$ is the geometrical frustration of Yb$^{3+}$ magnetic moments on the distorted Kagome lattice. It is anticipated that the metamagnetic transition is associated with a weakening of the magnetic frustration. Once the metamagnetic transition is tuned towards zero temperature, the long-wavelength critical fluctuations of the order parameter start showing quantum mechanical character and play an important role. The coexistence of magnetic frustration and quantum fluctuations may reinforce each other, giving rise to an increasing proportion of inelastic scattering of electrons. Close to the quantum critical region, strong inelastic scattering of electrons can then lead to inefficient heat transport compared to its counterpart charge transport, resulting in the violation of the WF law.

A further field-induced QCP in YbAgGe was proposed by Schmiedeshoff {\it et al.}  near 7.2 T, evidenced by the onset of a divergence of the Gr\"{u}neisen ratio \cite{Canfield2011}. This may account for the finite field region between 7 and 9 T, for which NFL behavior has been found. In the $T = 0$ K limit, however, the verification of WF law at $H = 7.2$~T implies that the quasiparticles still survive in this NFL region.

In summary, we have studied the low-temperature electrical resistivity and thermal conductivity of YbAgGe. The temperature dependence of the Lorenz ratio $L(T)/L_0$ indicates an anomalous suppression at the  critical field $H_c =4.5$~T of a metamagnetic quantum critical end point. Although we cannot neglect or subtract the phonon and magnon contributions to thermal conductivity, the observed linear temperature dependence of $L(T)/L_0$ for all different fields allows to extrapolate towards $T\rightarrow 0$. Such extrapolation yields a value significantly below 1 ($L/L_{0} = 0.92\pm 0.03$) only very close to $H_c$. This suggests strong inelastic scattering even at absolute zero temperature due to the disintegration of heavy quasiparticles. We speculate that magnetic frustration and the more local magnetic character compared to itinerant metamagnets like Sr$_3$Ru$_2$O$_7$ and CeRu$_2$Si$_2$ is responsible for this behavior.

We thank M.A. Tanatar for critical reading and useful suggestions. J.K. Dong acknowledges support from the Alexander-von-Humboldt Foundation. This work was supported by the German Science Foundation through the research unit 960 (Quantum Phase Transitions). Work done at Ames Laboratory (PCC and SLB) was supported by the U.S. Department of Energy, Office of Basic Energy Science, Division of Materials Sciences and Engineering. The Ames Laboratory is operated for the U.S. Department of Energy by Iowa State University under Contract No. DE-AC02-07CH11358.\\


\vspace{1cm}

\textbf{\em{Supplemental Material}}

\section{Experimental setup}

Single crystals of YbAgGe were grown from a AgGe-rich ternary
solution \cite{Morosan}. Our sample has a hexagonal cross
section, with $\sim$ 3 mm length parallel to the $c$-axis, along which the heat and charge transport has been realized. Contacts were made directly on the sample surfaces with silver epoxy and were used for both resistivity and thermal conductivity measurements. The contacts are metallic with typical resistances of less than 200 $m\Omega$ at room temperature, leading to a ratio of the contact to the sample resistance of about 3. However, this ratio should drop below 1 at low temperatures, because of the relatively small residual resistivity ratio of YbAgGe. It has been shown for silver epoxy contacts on high-purity CeCoIn$_5$, that the electron-phonon decoupling leads to a downturn of $\kappa(T)/T$ at very low temperatures~\cite{Tanatar1S}. Such behavior has been absent in our experiments. For the two neighboring voltage contacts, the ratio of each contact width over their distance is approximately 6$\%$, which can be taken as systematic standard error in our measurements. A dilution refrigerator with a superconducting magnet has been utilized, where the field has been applied along the $ab$ plane perpendicular to the electrical and heat current, respectively.  The electrical resistivity and thermal conductivity have been measured using home-made probes. For electrical resistivity, the base temperature has been 30~mK. For the thermal conductivity a standard four-wire steady-state method with two RuO$_2$ chip thermometers, calibrated
{\it in situ} against a reference RuO$_2$ thermometer located in a field-compensated zone has been utilized. The temperature gradient has been set to between 5\% and 7\% and the lowest accessible average temperature in these measurements has been 40~mK. Magnetic fields were applied along the $ab$ plane perpendicular to the electrical and heat current.

\section{Electrical resistivity measurements}

\begin{figure}[t!]
\includegraphics[clip,width=1\columnwidth]{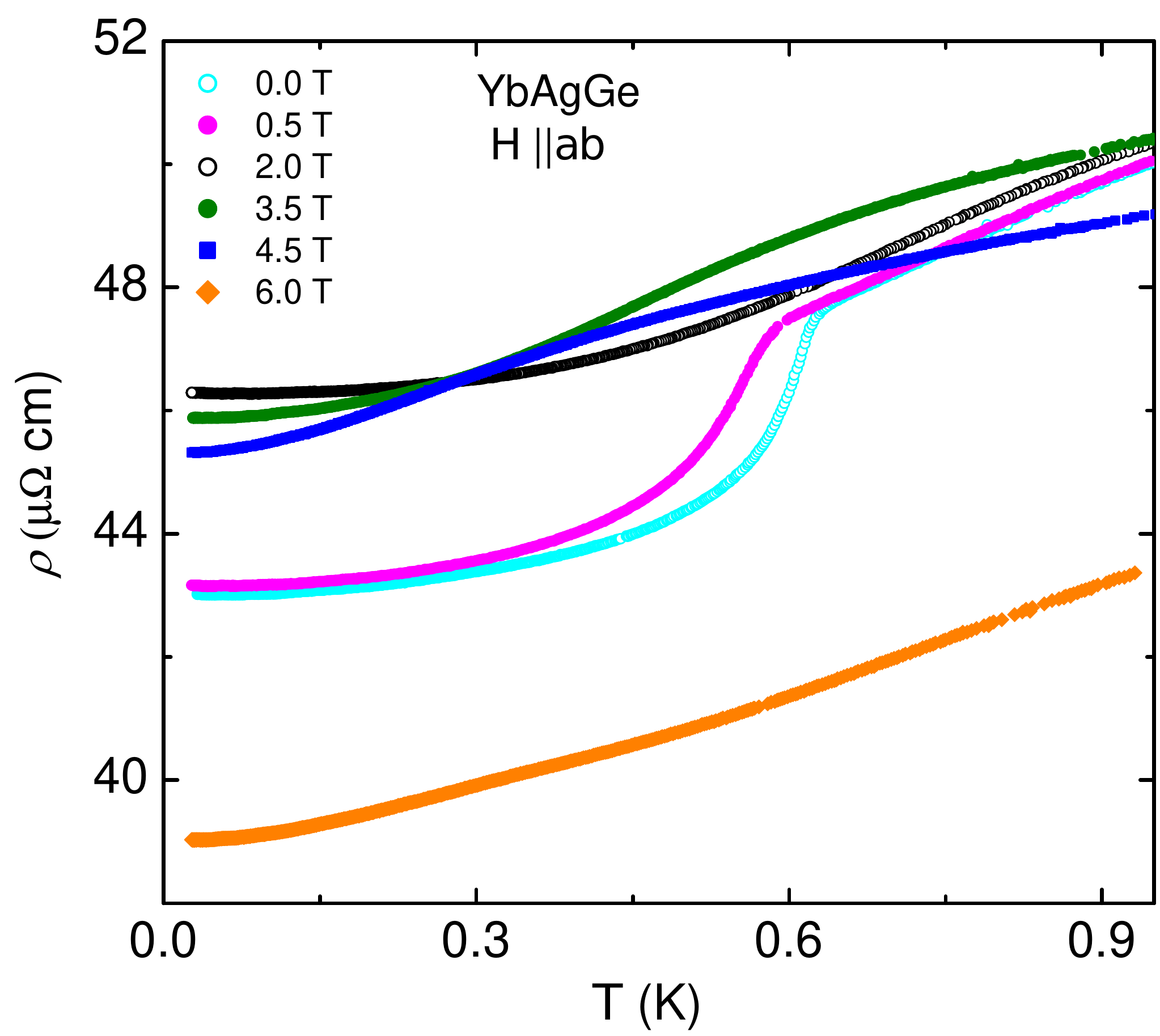}
\renewcommand{\thefigure}{S1}
\caption{Temperature dependence of the electrical resistivity $\rho(T)$ of YbAgGe along the $c$-direction at various magnetic fields $H \parallel ab$.}
\end{figure}

\begin{figure}[t!]
\includegraphics[clip,width=1\columnwidth]{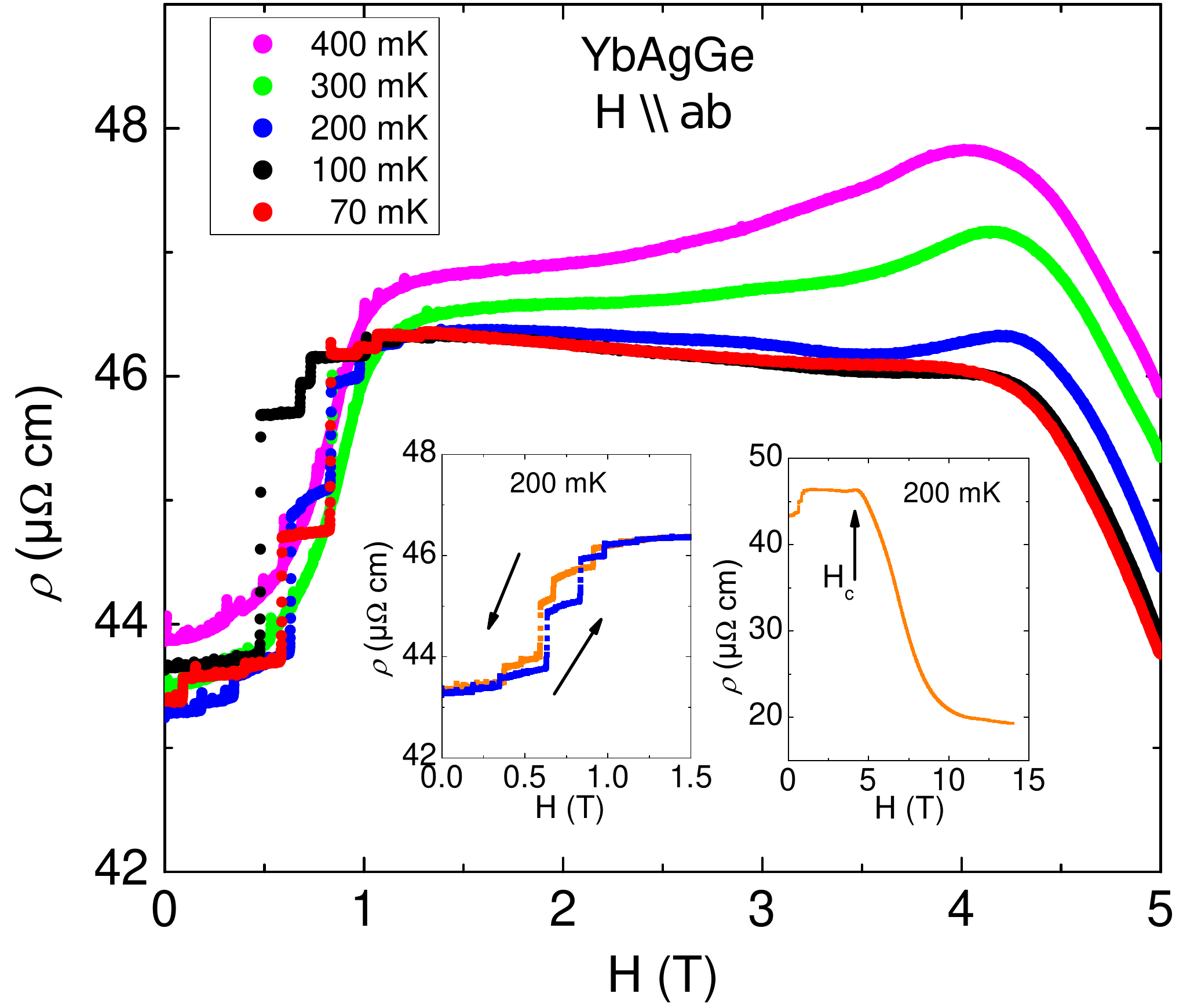}
\renewcommand{\thefigure}{S2}
\caption{(Color online). Field dependence of $\rho(H)$ isotherms for YbAgGe at constant temperatures between 70 and 400 mK. The left inset enlarges the low-field region for 200 mK. The arrows indicate the direction of the field sweeps. The right inset displays the 200 mK data at fields up to 14 T. The arrow indicates the critical field $H_c=4.5$~T.}
\end{figure}

Electrical resistivity measurements as function of temperature are shown Fig. 2 of the main part of the paper. In Fig. S1, we magnify the temperature dependence of the electrical resistivity for all the different investigated magnetic fields up to 6 T.

Fig. S2 shows the isothermal field dependence of resistivity
$\rho(H)$ for magnetic fields up to 5 T at various low temperatures. Two clear features can be seen. The first one appears at $H \simeq$ 1 T, the second in the vicinity of critical field ($\simeq$ 4.5 T), in agreement with the previous resistivity measurements measured down to 0.4 K \cite{Canfield2004S}. The resistivity anomaly at $H
\simeq$ 4.5 T may be related to a metamagnetic crossover between two different magnetically ordered phases~\cite{Canfield2004S}, although it remains continuous and broad down to 70~mK. Similar behavior has also been observed near the metamagnetic transition in CeRu$_2$Si$_2$~\cite{DaouS}, while for Sr$_3$Ru$_2$O$_7$ the formation of an electronic nematic phase leads to much sharper signatures in the isothermal magnetoresistance~\cite{Grigera01S}. At intermediate fields (1 T $ < H < 4$ T), the resistivity $\rho(H)$ shows a relatively weak field dependence. At high temperatures, $T= 400$~ mK, $\rho(H)$ increases slowly with increasing magnetic field and displays a broad maximum around the critical field, in agreement with earlier results \cite{Canfield2004S}. Upon decreasing the temperature, this maximum is suppressed and a negative magnetoresistance already at 4~T develops. In the low-field regime below 1 T, anomalous behavior is discovered at low temperatures. As can be seen in the left inset of Fig. S4, several sudden jumps are found in $\rho(H)$. While we do not know the origin of these jumps, the observed hysteresis points at some variation of micromagnetic domains. The latter may be related to the ordering in the $a$ phase, resulting from different equivalent directions of the propagation vector $k=(1/3, 0, 1/3)$~\cite{FakS} or alternatively to the first-order nature of the a-b transition at low temperatures~\cite{NiklowitzS}.

\section{Thermal conductivity measurements}

\begin{figure}[t!]
\includegraphics[width=1\columnwidth]{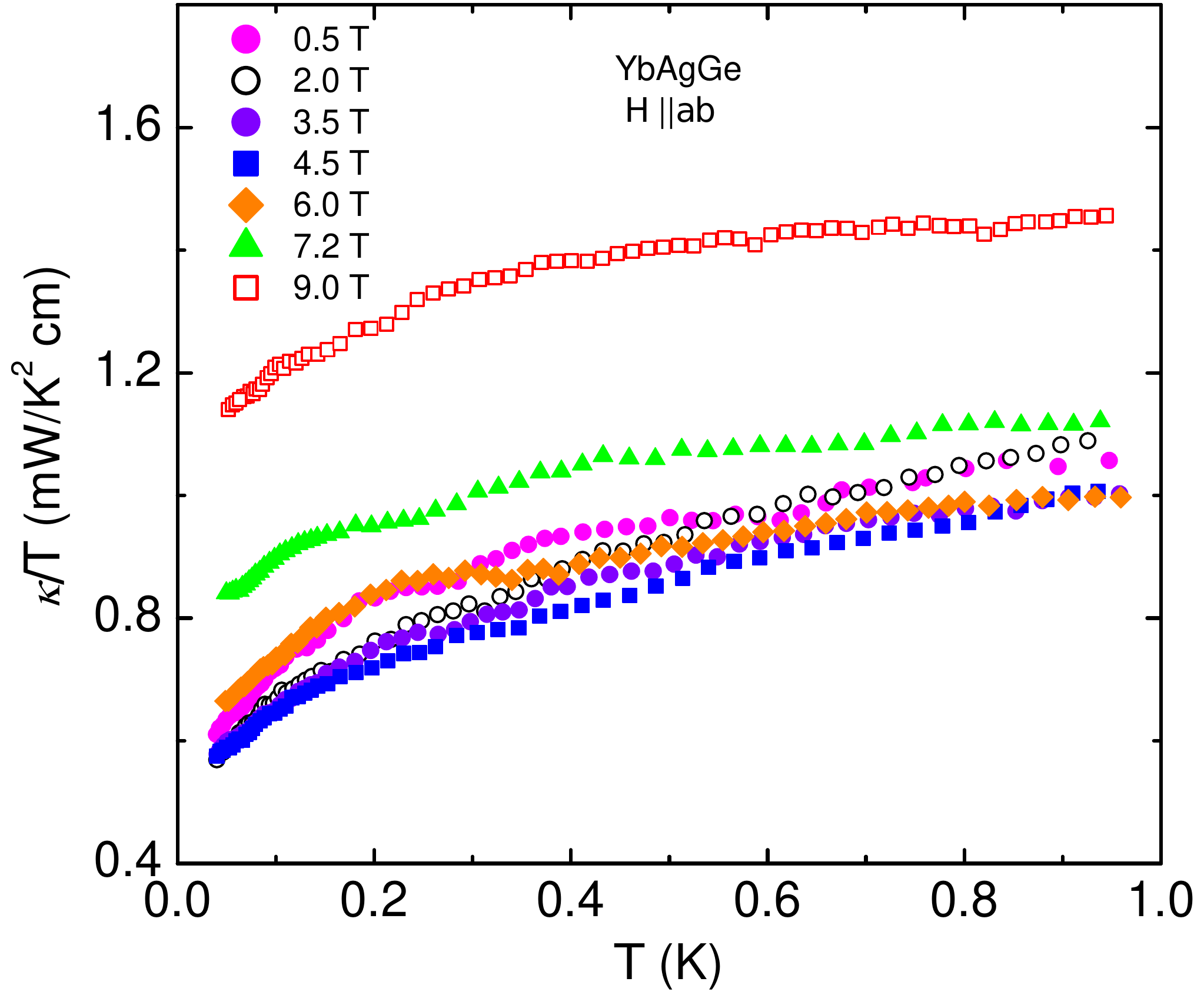}
\renewcommand{\thefigure}{S3}
\caption{Temperature dependence of the thermal conductivity as $\kappa/T$ vs. $T$ of YbAgGe along the $c$-direction at various magnetic fields $H \parallel ab$.}
\end{figure}

In the main part of the paper, the thermal conductivity results are displayed for a selection of applied magnetic fields and at temperatures below 0.3 K. In Fig. S3, the full data sets up to 0.95~K are shown, at all different applied fields.

\section{Analysis of the Lorenz ratio}

\begin{figure}[t!]
\includegraphics[width=1\columnwidth]{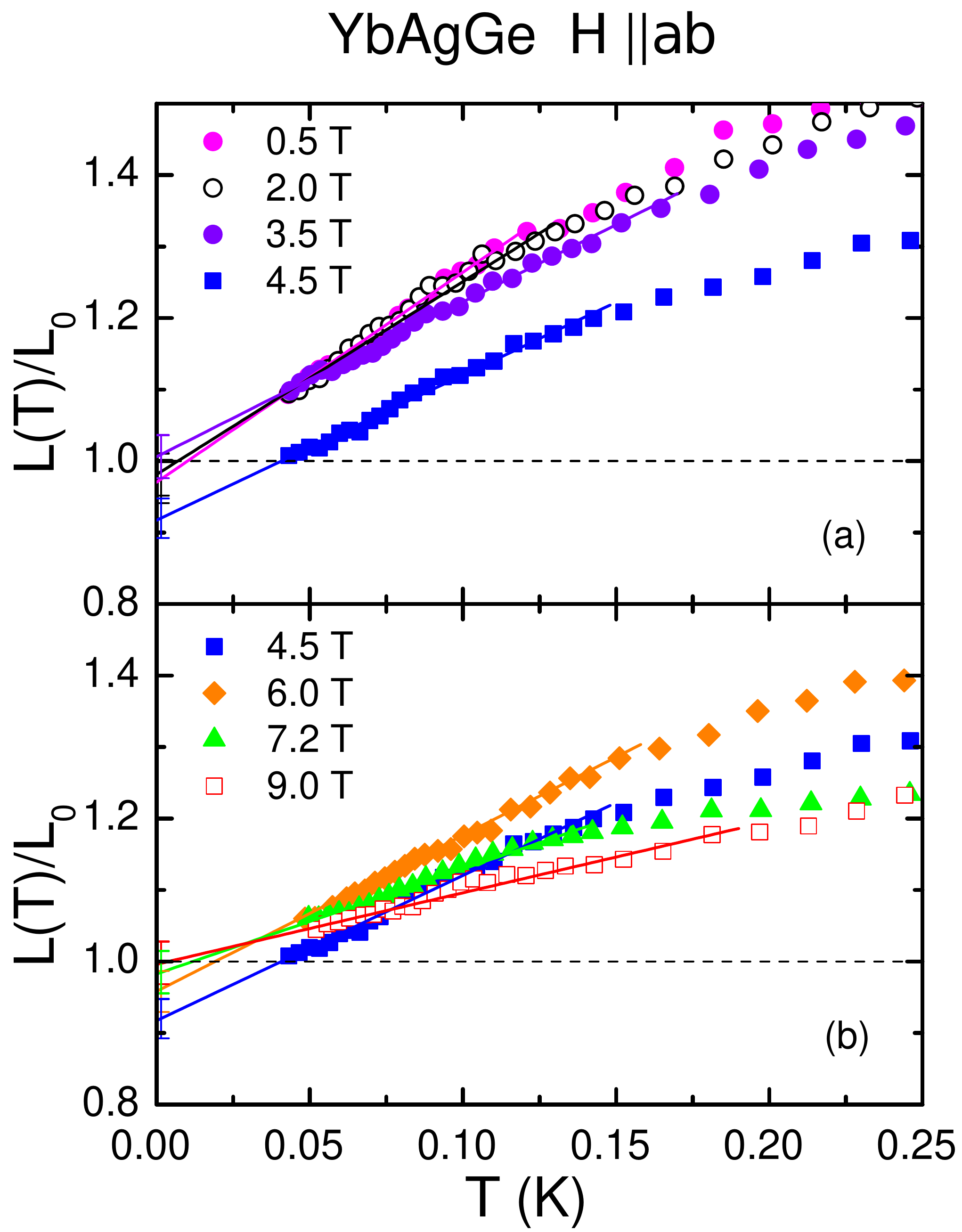}
\renewcommand{\thefigure}{S4}
\caption{Temperature dependence of the normalized Lorenz ratio $L(T)/L_{0}$ = $\kappa\rho/(L_{0}T)$ for all different applied magnetic fields. The solid lines indicate linear fits to the data. Error bars at data points represent the statistical scattering while those at the lines near absolute zero temperature indicate the systematic error for the extrapolated Lorenz ratio arising from the finite width of the contacts.}
\end{figure}

Fig. S4 displays the temperature dependence of the normalized Lorenz ratio $L(T)/L_0=(\kappa/T)\cdot \rho/L_0$ with $L_0=(\pi^2/3)\cdot (k_B/e)^2$. The plot displays all measured data at temperatures below 0.25, where data for fields up to 4.5 T are shown in part a and data from 4.5 T up to 9 T are shown in part b. The lines display linear fits from the lowest measured temperature up to a variable maximal temperature for which $T$-linear behavior is found. A significant deviation of the WF law is observed at the critical field of 4.5~T, for which the extrapolation reveals $L/L_0=0.92\pm 0.03$. In comparison to all other investigated fields, the data at the critical field indicate a pronounced depression of the Lorenz ratio $L(T)$ near 50 mK.

\end{document}